\documentclass[amsmath,amssymb,12pt]{article}
\usepackage[all]{xy}
\usepackage[centertags]{amsmath}
\usepackage{latexsym}
\usepackage{amsfonts}
\usepackage{amssymb}
\usepackage{amsthm}
\usepackage{graphicx}
\def\R{{\rm I\!R}}

\begin{document}
\title{Focusing: coming to the point in metamaterials}
\author{S. Guenneau$^{a,b}$$^{\ast}$, A. Diatta$^{b}$ and R.C. McPhedran$^{c}$ }
\maketitle
{\footnotesize
\thanks{
\noindent
$^{a}${\em{Institut Fresnel, UMR CNRS
6133, University of Aix-Marseille,
\newline\noindent
case 162, F13397 Marseille Cedex 20, France
\newline\noindent
Email: guenneau@liv.ac.uk}};
\\
$^{b}${\em{Department of Mathematical
Sciences, Liverpool L693BX, UK
\newline\noindent
Email: adiatta@liv.ac.uk}};
\\
$^{c}${\em{IPOS, School of Physics, University of Sydney NSW,
Australia
\newline\noindent
Email: ross@physics.usyd.edu.au}
}
}

\begin{abstract}
This paper reviews some properties of lenses in curved and folded
optical spaces. The point of the paper is to show some limitations
of geometrical optics in the analysis of subwavelength focusing. We
first provide a comprehensive derivation for the equation of
geodesics in curved optical spaces, which is a tool of choice to
design metamaterials in transformation optics. We then analyse the
resolution of the image of a line source radiating in the Maxwell
fisheye and the Veselago-Pendry slab lens. The former optical medium
is deduced from the stereographic projection of a virtual sphere and
displays a heterogeneous refractive index $n(r)$ which is
proportional to the inverse of $1+r^2$. The latter is described by a
homogeneous, but negative, refractive index. It has been suggested
that the fisheye makes a perfect lens without negative refraction
[Leonhardt, Philbin arxiv:0805.4778v2]. However, we point out that
the definition of super-resolution in such a heterogeneous medium
should be computed with respect to the wavelength in a homogenized
medium, and it is perhaps more adequate to talk about a conjugate
image rather than a perfect image (the former does not necessarily
contains the evanescent components of the source). We numerically
find that both the Maxwell fisheye and a thick silver slab lens lead
to a resolution close to $\lambda/3$ in transverse magnetic
polarization (electric field pointing orthogonal to the plane). We
note a shift of the image plane in the latter lens. We also observe
that two sources lead to multiple secondary images in the former
lens, as confirmed from light rays travelling along geodesics of the
virtual sphere. We further observe resolutions ranging from
$\lambda/2$ to nearly $\lambda/4$ for magnetic dipoles of varying
orientations of dipole moments within the fisheye in transverse
electric polarization (magnetic field pointing orthogonal to the
plane). Finally, we analyse the Eaton lens for which the source and
its image are either located within a unit disc of air, or within a
corona $1<r<2$ with refractive index $n(r)=\sqrt{2/r-1}$. In both
cases, the image resolution is about $\lambda/2$.
\bigskip

 \bigskip

\end{abstract}

\section{Introduction}
In 1967, the Russian physicist Victor Veselago wrote a visionary
paper in which materials with a negative refractive index were
theorized \cite{veselago}. Veselago pointed out that this could
happen only if the real parts of both dielectric permittivity
$\epsilon$ and magnetic permeability $\mu$ are simultaneously
negative at a given frequency $\omega$. It was argued by a ray
analysis that a  slab of such a negative index material (NIM) can
act as a flat lens that imaged a source on one side to a point on
the other. But this result remained an academic curiosity for almost
thirty years, until Sir John Pendry and co-workers
\cite{pendry96,pendry_IEEE} proposed designs of structured materials
which would have effectively negative $\varepsilon$ and $\mu$.
However, this very unlikely event occurs only in a very narrow range
of frequencies and in real life, NIM are necessarily dissipative and
dispersive. Interestingly, one of us studied together with Graeme
Milton and Nicolae-Alexandru Nicorovici the electrostatic response
of a coated cylinder with negative $\epsilon$ back in 1994
\cite{Ross94}, and this can be also considered as a perfect lens in
the intense near field limit in transverse electric polarization
\cite{milton1,milton2,milton3}.

In 2000, the experimental demonstration of a negative refractive
index at GHz frequencies by a team led by David Smith \cite{smith00}
provided a fillip to research in this area (see
\cite{pendry_contemp04,sar_rpp05} for recent reviews). One should
note that these metamaterials are structured at subwavelength
lengthscales (typically one tenth of the wavelength) and it is
possible to regard them as homogeneous and describe their response
by dispersive effective medium parameters (see \cite{miltonbook} for
a comprehensive survey on homogenization theory). Potential
applications of negative refraction came when Pendry demonstrated
that the Veselago slab lens not only involves the propagation waves
but also the evanescent near-field components of a source in the
image formation\cite{pendry_prl00}. However, there is still some
limitation to the resolution (which is hardly surprising given the
laws of physics), which can only become infinite in the limit of
zero absorption, even though it is possible to improve the lens
resolution by considering multilayered negatively refracting lenses
or by adding some gain \cite{sar1,sar2,sar3}.

In 2006, the physicists John Pendry, David Schurig and David Smith
theorized that a finite size object surrounded by a spherical
coating consisting of a metamaterial might become invisible for
electromagnetic waves \cite{pendrycloak}. This proposal has been
experimentally validated the same year in the microwave regime using
a two-dimensional configuration \cite{cloakex}. The same year, Ulf
Leonhardt independently proposed a conformal map route towards
cloaking \cite{ulf2006,ulf0}, which is valid in the geometrical
optics limit (when the wavelength is much smaller than the
diffracting obstacle). Both proposals actually derive from the
earlier study by Allan Greenleaf, Yaroslav Kurylev, Matti Lassas,
and Gunther Uhlmann \cite{greenleaf} whereby the conductivity of an
object was much reduced.

However, an alternative route towards cloaking using anomalous
resonances in negatively refracting cylindrical lenses based upon
the earlier work by McPhedran, Nicorovici and Milton \cite{milton2}
in 1994, shows that transformation optics and perfect lensing are
intimately linked. A perfect lens can be seen as a multi-valued map
whereby a source is mapped twice onto itself (a NIM is in essence a
folded optical space) as first shown by Ulf Leonhardt and Thomas
Philbin \cite{Ulf} and then further investigated by Milton et al.
\cite{milton3} and subsequently by one of us with S. Anantha
Ramakrishna \cite{sgsar09}. Using such transformation optics tools,
it was also shown that two corners of NIM combined in a checkerboard
fashion can act as a unique resonator
\cite{pendry_jpc03,guenneau_ol05,notomi,he_njp}. Such checkerboards
and can be themselves mapped onto three dimensional corner
reflectors \cite{guenneau_njp05} and they actually exhibit some form
of extraordinary transmission \cite{sar_guenneau_OE2006}.

Furthermore, in 2008 Ulf Leonhardt and Thomas Tyc proposed an
improved type of cloaking \cite{tyc1}, based upon a stereographic
projection of a sphere onto a flat plane, leading to a non singular
nearly perfect cloak. A stereographic projection is a particular
mapping that projects a sphere onto a plane. The projection is
defined on the entire sphere, except at one point, the projection
point. Where it is defined, the mapping is smooth and bijective. It
is conformal, meaning that it preserves angles. It is neither an
isometry nor area-preserving: that is, it preserves neither
distances nor the areas of figures. Using such a conformal mapping,
Leonhardt and Tyc further designed some super antennas
\cite{tyc2,tyc3}. However, the materials resulting from these
optical transformations are highly heterogeneous, and it is
therefore legitimate to ask whether they fall within  a class of
super resolution optical systems and indeed specify in which sense
they could be considered as high-resolution devices. This question
is of foremost importance as some researchers have recently shown
that simple enough tomography devices beat the diffraction limit
\cite{hughes}, however with the constraint that the source be
located close to a structured surface (a grating) and therefore
perfect lensing occurs only in the near field limit.

The plan of the paper is as follows: following this first
introductory section, we discuss in section 2 the variational
framework associated with Fermat's principle, also known as
principle of least time, which is the idea that the path taken
between two points by a ray of light is the path that can be
traversed in the least time. Section 3 is then devoted to the design
of the Maxwell fisheye and the Veselago-Pendry lens through
transformation optics. Section 4, the core of the paper, addresses
the issue of whether the fisheye can be considered as a perfect lens
without negative refraction. We look both at the cases of a fisheye
of infinite extent and a fisheye surrounded by perfect conducting
boundaries, as first introduced by Ulf Leonhardt in 2009
\cite{fishulf}. Section 5 discusses the issue of multiple (mirage)
images within the fisheye when there are two or more sources.
Section 6 is devoted to the analysis of the Eaton lens, which shares
common some features with the mirror fisheye. Finally, section 7
draws concluding remarks.

\begin{figure}[h!]\centerline{
\includegraphics[width=12cm]{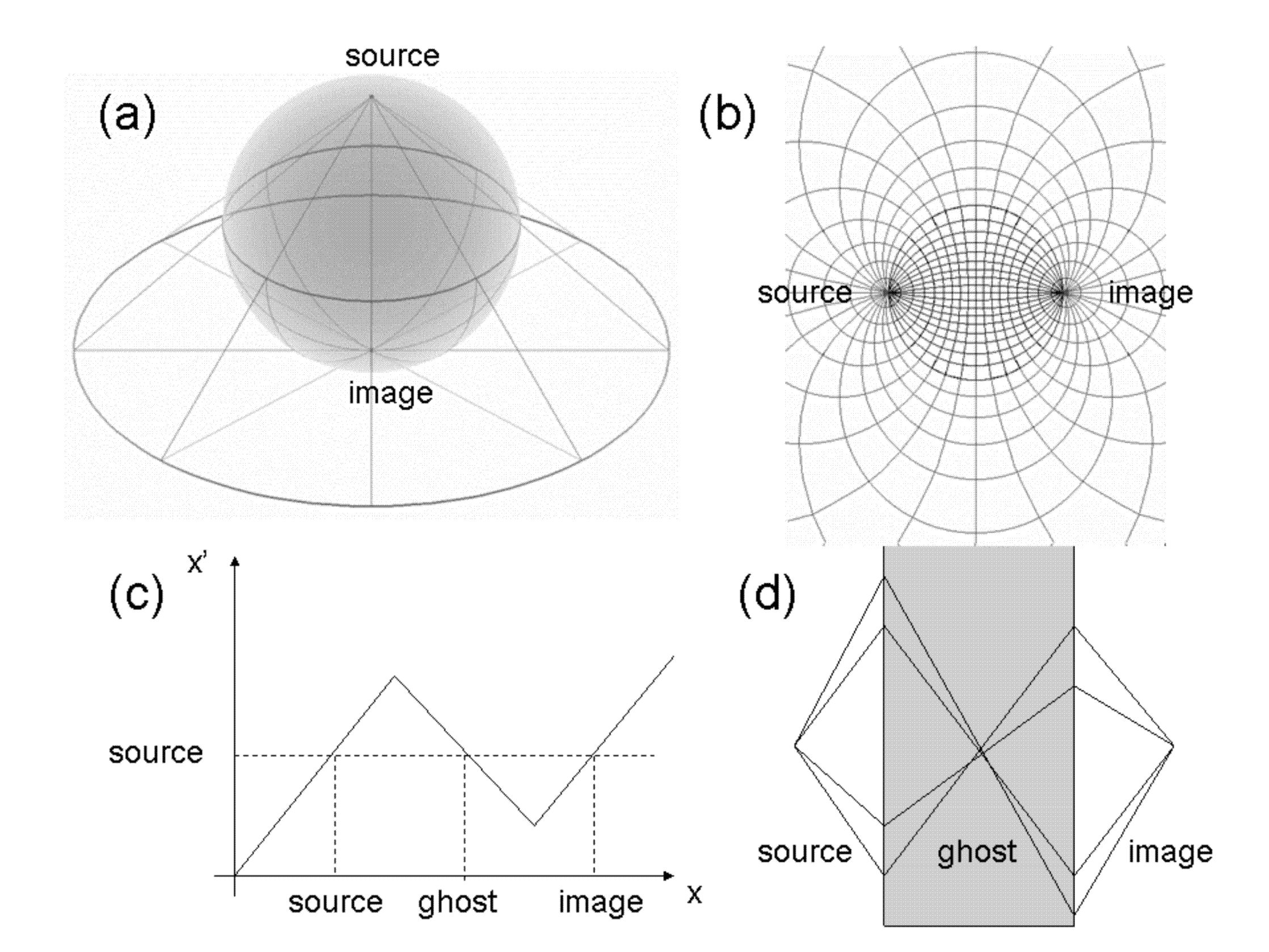}}
\caption{\em\small ~Upper panel: The Maxwell fish eye; (a) Stereographic
projection of a virtual sphere onto a physical plane; A source has
an antipodal image on a great circle (or ray trajectory); (b)
Geodesics (or ray trajectories) between a source and its image on
the physical plane. Lower panel: The Veselago-Pendry slab lens
(adapted from \cite{ulf1}); (c) Coordinate transformation from the
physical $x$-axis to the virtual axis $x'$; The inverse
transformation from $x'$ to $x$ is either triple- or single-valued.
The triple-valued segment on the physical $x$-axis corresponds to
the focal region of the lens; (d) A source point has two images on
the physical axis i.e. one inside the lens (a ghost image) and one
on the other side (a perfect image).} \label{fig0}
\end{figure}

\section{Geometry of geodesics on a sphere}
Let us consider the sphere $S^2=\{(x',y',z')\in \mathbb R^3 \text{
with } x'^2+y'^2+z'^2=r_0^2\}$ of radius $r_0$, as shown in Fig.
\ref{fig0}(a). Ray trajectories follow so-called geodesics (great
circles) on this sphere i.e. shortest trajectories. A geodesic
$x^i$, for a metric $g_{ij}$ is a curve $t\mapsto x^i(t)$ which is a
minimum of the integral path
\begin{equation}
\int_{a}^{b} \, ds =
\int_{a}^{b}\sqrt{g_{ij}\frac{dx^i(t)}{dt}\frac{dx^j(t)}{dt}} \,
dt:= \int_{a}^{b}\sqrt{g_{ij}\dot{x}^i\dot{x}^j} \, dt
\end{equation}
between two points $a$ and $b$ in a curved space.

In transformation optics, the principle of least action of Pierre de
Fermat is often invoked to deduce that the eikonal equation (linking
the electric field intensity to the gradient electric potential)
which describes the phase front of waves (in the ray optics limit)
admits a geodesic as a local solution. Unfortunately, a global
solution e.g. a solution for all time in the geometrical optics case
is not possible. The reason is that caustics may develop which means
that wavefronts cross. When the ray optics picture breaks down, it
is then necessary to solve the vector Maxwell equations which are
always valid in a linear context.

For the sake of completeness, let us establish with basic arguments
that the equation of geodesics reads
\begin{equation}
g_{ij}\ddot{x}^j(s)+\frac{1}{2}g_{lj,i}\dot{x}^l(s)\dot{x}^j(s)=0 \;
, \label{maingeo}
\end{equation}
where $g_{lj,i}$ denotes the derivative of $g_{lj}$ with respect to
$x^i$.

In classical optics, the eikonal equation is known to be valid when
the wavelength $\lambda$ is small in comparison to the size of the
diffracting obstacle. However, in highly-heterogeneous media such as
metamaterials deduced from non-Euclidean transformations, one should
also take into account the effect of the optical space curvature
which should be small on scales compatible with wavelengths, that is
$\mid R \mid \ll \omega^2/c^2$, where $R$ is the scalar curvature
$R=g^{ij}R_{ij}$ \cite{tyc1}: local space curvature should not be on
the same scale as the electromagnetic wave oscillations. We note
that in the case of the sphere which has obviously non zero
curvature, the eikonal equation is a good approximation to describe
trajectories of light only when the wavelength is much smaller than
the radius of the sphere.

To establish (\ref{maingeo}), let us consider the classical
minimization problem:
$${\cal P}: \inf_{{\bf x}\in C^1([a,b],S^2)}\left \{ E({\bf x})
=\int_{a}^b L(t,{\bf x}(t),\dot{{\bf x}}(t)) \, dt \; ; \; {\bf
x}(a)=\alpha \; , \; {\bf x}(b)=\beta \right\}$$

\noindent where $L=g_{ij}\dot{x}^i\dot{x}^j\in C^2([a,b]\times
S^2\times\R^3,\R)$ {\it i.e.} $L$ is a continuous function with
continuous second derivatives such that $L:(t,v,\xi)\in [a,b]\times
S^2\times\R^3 \to L(t,v,\xi)\in \R$.

\noindent For ${\bf \alpha}$ and ${\bf\beta}$ close enough in $S^2,$
let ${\bf x}\in C^2([a,b], S^2)$ be a minimum of $E$ in the convex
set $K$:
$$K_{\bf \alpha,\beta}=\{{\bf v}\in C^1([a,b],S^2)
\; , \; {\bf v}(a)=\alpha\; , \; {\bf v}(b)=\beta\}$$

\noindent Then, we can write that for every $\phi\in
C^1([a,b],\R^3)$ with $\phi(a)=\phi(b)={\bf 0}$, and for every
$\theta\in]-1,1[$ such that ${\bf x}+\theta\phi\in S^2$
\footnote{Take e.g. $\phi$ such that ${\bf x}(t)
+\frac{\theta}{2}\phi(t)$ is orthogonal to $\phi(t)$ and
$\phi(a)=\phi(b)={\bf 0}$.}
$$E({\bf x}) \leq E({\bf x}+\theta\phi) \; .$$

\noindent In particular, we are assured that \footnote{We use the
fact that ${\bf x}\in C^2([a,b],\R^3)$ since differentiating $E$
involves differentiating the integral of $L(t,{\bf x},\dot{{\bf
x}})$ over $[a,b]$ which requires $\ddot{{\bf x}}$ to be continuous
on $[a,b]$.}
$${\left\{\frac{d}{d\theta} E({\bf x}+\theta\phi)\right\}}_{\mid\theta=0}=0 \; .$$

This leads us to
$${\left\{\frac{d}{d\theta} \int_a^b L(t,{\bf x}(t)+\theta\phi(t),\dot{{\bf x}}(t)+\theta\dot{\phi}(t))
\, dt \right\}}_{\mid\theta=0}=0 \; .$$

Thus, we obtain
$$\int_a^b \left\{ \frac{\partial L}{\partial v}\phi+\frac{\partial L}{\partial
\xi}\dot{\phi} \right\} \, dt =0 \; .$$

\noindent This equation holds for any $\phi\in C^1([a,b],\R^3)$ such
that $\phi(a)=\phi(b)$ and $\theta\phi\in S^2$, hence integrating by
parts we have
$$\int_a^b \left\{ \frac{\partial L}{\partial v^i}-\frac{d}{dt}\frac{\partial L}{\partial
\xi^i}\right\} \phi \, dt =0 \; .$$

\noindent Applying the fundamental lemma of variational calculus
\cite{daco}, we are assured that for every $t\in(a,b)$,
$$\frac{d}{dt}\left[ \frac{\partial}{\partial\xi} L(t,{\bf x}(t),\dot{{\bf x}}(t))\right]
=\frac{\partial}{\partial v} L(t,{\bf x}(t),\dot{{\bf x}}(t)) \; .$$

\noindent The geodesics are thus given as solutions of the
Euler-Lagrange equations, which take the following form:
$$\frac{d}{dt}\left[ g_{ij}\dot{x}^j(t)\right]
=g_{lj,i}\dot{x}^l(t)\dot{x}^j(t) \; .$$

\noindent This leads us to (\ref{maingeo})
$$g_{ij}\frac{d^2
x^i(s)}{ds^2}+g_{lj,i}\frac{dx^l(s)}{ds}\frac{dx^j(s)}{ds}
=\frac{1}{2} g_{lj,i}\frac{dx^l(s)}{ds}\frac{dx^j(s)}{ds} \; ,$$ and
may be further simplified \footnote{Indeed, by multiplying this
equation by the inverse tensor $g^{ij}$ from the left and
introducing the Christoffel symbol
$\Gamma^i_{jk}=\frac{1}{2}g^{il}(g_{lj,k}+g_{lk,j}-g_{jk,l})$, we
obtain $\displaystyle \ddot{x}^k(s)+\Gamma_{ij}^k(s)
\dot{x}^i(s)\dot{x}^j(s)=0$, where the $\Gamma_{ij}^k$'s are the
Christoffel symbols of the Levi-Civita connection of $g.$}.

\noindent Using the expression
\begin{equation}
ds^2=r_0^2(d\theta^2+\sin^2\theta d\phi^2) \; ,
\end{equation}
for the metric on the surface of a sphere of radius $r_0$, it can be
checked that geodesics on a sphere are nothing but the great
circles. We note that this expression cannot be reduced to the
Euclidean form $ds^2={(dx^1)}^2+{(dx^2)}^2$, which shows that the
surface of a sphere is not a Euclidean space (it has non-zero
curvature).

\section{Design of the Maxwell fisheye and the Veselago-Pendry lens through transformation optics}

In this section, we use the conventional notation
$\{x,y,z\}=\{x^1,x^2,x^3\}$ for the Euclidean system of coordinates.
Whereas the derivation of the refractive index within the Maxwell
fish eye can be found in the literature (see \cite{ulf1} for a
comprehensive review), we include it for completeness.

Let us consider the sphere $S^2=\{(x',y',z')\in \mathbb R^3 \text{
with } x'^2+y'^2+z'^2=r_0^2\}$ of radius $r_0$, as shown in Fig.
\ref{fig0}(a). A line element on the virtual sphere of radius $r_0$
is given by:
\begin{equation}
ds^2=dx'^2+dy'^2+dz'^2 \; , \hbox{ with } \; x'^2+y'^2+z'^2=r_0^2 \; .
\label{infelement}
\end{equation}

Using the stereographic projection:
$$x=\frac{x'}{1-z'/r_0} \; , \; y=\frac{y'}{1-z'/r_0} \; ,$$
together with the inverse formula
 \begin{eqnarray}\label{stereo-inv} x'=\frac{2x}{1+(r/r_0)^2}
 \; , \; y'=\frac{2y}{1+(r/r_0)^2} \; , \; z'= r_0\frac{(r/r_0)^2-1}{1+(r/r_0)^2}
\end{eqnarray}
where $ r^2=x^2+y^2$, and substituting in expression
(\ref{infelement}),
we end up with the expression for a line element in the projected
plane:
\begin{eqnarray}
ds^2=Pdx^2+ Q dxdy+R dy^2
\end{eqnarray}
where
\begin{eqnarray*}
&P=\displaystyle{\left(\frac{\partial x'}{\partial
x}\right)^2+\left(\frac{\partial y'}{\partial x}\right)^2+\left(\frac{\partial z'}{\partial
x}\right)^2}, \nonumber \\
&Q= \displaystyle{\frac{\partial x'}{\partial x}\frac{\partial x'}{\partial
y} +\frac{\partial y'}{\partial x}\frac{\partial y'}{\partial
y}+\frac{\partial z'}{\partial x}\frac{\partial z'}{\partial y}} \; , \;
R=\displaystyle{\left(\frac{\partial x'}{\partial y}\right)^2+\left(\frac{\partial y'}{\partial
y}\right)^2+\left(\frac{\partial z'}{\partial y}\right)^2} \; .
\end{eqnarray*}
In our particular case of the sphere and using the above expression
(\ref{stereo-inv}) of the inverse of  the stereographic projection,
we get $Q=0$ and $P=Q$, so that the corresponding line element on
the projected plane simplifies into:
\begin{equation}
ds^2=n^2(dx^2+dy^2) \; , \; n= \frac{2 r_0^2}{x^2+y^2+r_0^2} \; .
\label{infelementplane}
\end{equation}

Let us now apply these mathematical tools to the slab perfect lens
which is interesting inter alia since it corresponds to a
non-one-to-one coordinate transformation. This is clear, since it
has triplets of planes on which the field distributions are
identical: the object plane, the internal image plane and the
external image plane. In the one-dimensional case, the corresponding
coordinate transformation maps these three planes from a single
plane in the reference space. Consider the coordinate transformation
of Fig. \ref{fig0}(c) which is given by
\begin{equation}
x' = x -d \; , \hbox{ if } x'< d/2 \hbox{, or } -x\hbox{ if } -d/2 <
x' < d/2 \hbox{, or } x +d \hbox{ if } d/2 < x'
\label{transfolens1d}
\end{equation}
where d is the thickness of the lens.

This leads to the identity transformation outside the lens, whereas
inside the lens i.e. for $-d/2 < x' < d/2$, the derivative of
$x(x')$ becomes negative as $dx/dx'=-1$ which flips the sign of $n$.
Moreover, there is no change in $y$ and $z$ coordinates, so that the
material properties differ from free space only in the $x=x'$
direction, with $n=-1$ inside the lens and $n=+1$ outside. However,
$dx/dx'$ is undefined at $x'=\pm d/2$ and so is $n$ at these
interfaces.

\section{A perfect lens without negative refraction?}
As pointed out in the abstract, there might be an ambiguity in the
definition of a perfect image in metamaterials, which are highly
heterogeneous structures. We therefore decided to address this issue
using full wave computations Thanks to the cylindrical geometry, the
problem splits into TM and TE polarizations:
\begin{equation}
\nabla\times\left( \nabla \times{\bf E}_l \right) -
\mu_0\varepsilon_0\omega^2 n^2{\bf E}_l=-i\omega I_s \mu_0
\delta_{\mathbf{r}_s} {\bf e}_z  \; , \label{tm}
\end{equation}
\begin{equation}
\nabla\times\left( n^{-2}\nabla\times {\bf H}_l \right) -
\mu_0\varepsilon_0\omega^2 {\bf H}_l= \nabla\times (n^{-2}
\mathbf{j}_T ),
\label{te}
\end{equation}
where ${\bf E}_l=E_3(x,y){\bf e}_z$, ${\bf H}_l=H_3(x,y){\bf e}_z$
$\varepsilon_0\mu_0=1/c^2$, with $c$ the celerity of light in vacuum,
and $\omega$ the wave frequency. Here, the refractive index $n=n(x,y)$ is defined by
(\ref{infelementplane}) in the case of the Maxwell fisheye.
Moreover, $\mathbf{j}_T$ in (\ref{te}) denotes a current with a
vanishing $z-$component.

We note that in the TE case, the right-hand side of (\ref{te}) shows
that the very definition of the source, for instance a magnetic
dipole generated by a current circulating on a closed loop, depends
upon the surrounding medium. In a heterogeneous medium, the field
radiated by the source appears to be deformed by the spatially
varying refractive index $n$. To avoid additional numerical
technicalities in the finite element implementation and to simplify
the physical discussion, we focus on the TM case in the sequel,
except in subsection 4.3, in order to foster numerical efforts
towards the complete modelling of TE polarisation.

\subsection{Silver slab lens versus Maxwell's fisheye in unbounded domains}
The unbounded domain is modelled using perfectly matched layers
(PMLs), which are reflectionless heterogeneous anisotropic media
introduced by Berenger fifteen years ago \cite{berenger}. Nowadays,
in the time harmonic case, the most natural way to introduce PMLs is
to consider them as maps on a complex space
\cite{lassas,compel2008a} so that the corresponding change of
(complex) coordinates leads to equivalent $\varepsilon$ and $\mu$
(that are complex, anisotropic, and inhomogeneous even if the
original ones were real, isotropic, and homogeneous). This leads
automatically to an equivalent medium with the same impedance as the
one of the initial ambient medium since $\varepsilon$ and $\mu$ are
transformed in the same way and this insures that the interface with
the layer is non-reflecting at all incidences.

\noindent In the present study, the cylindrical PML is an annulus
whose characteristics are associated with the complex matrix
\cite{compel2008a}
\begin{equation}
\mathbf{T}_{PML}^{-1}=
\mathbf{R}(\theta)\mathbf{diag}(\frac{\tilde{\rho}}{s_\rho\rho},\frac{s_\rho\rho}{\tilde{\rho}},
\frac{s_\rho\tilde{\rho}}{\rho})\mathbf{R}(-\theta) \; ,
\end{equation}
with $\mathbf{R}(\theta)$ the rotation matrix by an angle $\theta$
in the $xy$-plane. This expression is the metric tensor in Cartesian
coordinates $(x,y,z)$ for the cylindrical PML. $\theta, \rho,
\tilde{\rho},$ and $s_\rho(\rho)$ are explicit functions of the
variables $x$ and $y$, $i.e.$ $\theta=2 \arctan\left(
\frac{y}{x+\sqrt{x^2+y^2}}\right)$, $\rho=\sqrt{x^2+y^2}$,
$s_\rho(\rho)=s_\rho(\sqrt{x^2+y^2})$,  and $
\tilde{\rho}=\int_0^{\sqrt{x^2+y^2}} s_\rho(\rho') d\rho'$ where
$s_\rho(\rho')$ is an arbitrary but well chosen complex valued
function of a real variable that describes the radial stretch
relating the initial radial distance $\rho$ to the complex one
$\tilde{\rho}$.

\noindent In this annulus, the governing equation (\ref{tm}) takes
the following form:
\begin{equation}
\nabla\times\left(\mathbf{T}_{PML}^{-1}\nabla \times{\bf E}_l
\right) - \mu_0\varepsilon_0\omega^2 n^2\mathbf{T}_{PML}^{-1}{\bf
E}_l={\bf 0} \; . \label{tmpml}
\end{equation}

We first consider the case of a refractive index with opposite signs
in a lens and the surrounding medium in (\ref{tm}), with $n$ set to
the positive value $1$ in the governing equation (\ref{tmpml}) for
the annulus. To preserve the (hypo) ellipticity of the formulation,
it is necessary to consider a small positive imaginary part to $n$
in (\ref{tm}) for the negatively refracting medium (e.g. modelling
absorption in metal for visible wavelengths). We thus consider $n=1$
(air) in the surrounding medium and $n=-1+i*0.4$ (silver) in the
lens. When the image is formed in the same homogeneous medium as the
source such as it is the case of such a Veselago-Pendry slab lens,
one shall compute the square modulus of the field in the image plane
and take the full-width at half maximum of the profile to compute
the resolution of the image, see Figs. \ref{fig6}(a) and
\ref{fig5}(a) for the case of a silver slab lens displaying a
resolution of $\lambda/3$. Interestingly, the image forms in an
image plane shifted by half a wavelength (equal to half the width of
the lens) compared with the theoretical prediction of the geometric
transform (\ref{transfolens1d}). We attribute this to the imaginary
part of the refractive index, which is in fact both spoiling the
image resolution (through absorption) and the mirror effect about
the mid-axis of the lens. We note that Pendry's poor-man lens was
experimentally shown to display a resolution of $\lambda/6$ by
Zhang's team back in 2005 \cite{zhang_science05}. However, the
source was located in the close neighborhood of a thin film of
silver i.e. in the intense near field limit. Here the image forms at
a finite distance from the lens, yet not exactly according to the
inverted Snell-Descartes laws of refraction. We numerically checked
that when we reduce the imaginary part of the negative refractive
index, the image forms closer to the predicted image plane, and the
resolution increases (the full-width at half maximum of the profile
narrows).

\begin{figure}[h!]
\centerline{
\includegraphics[width=10cm]{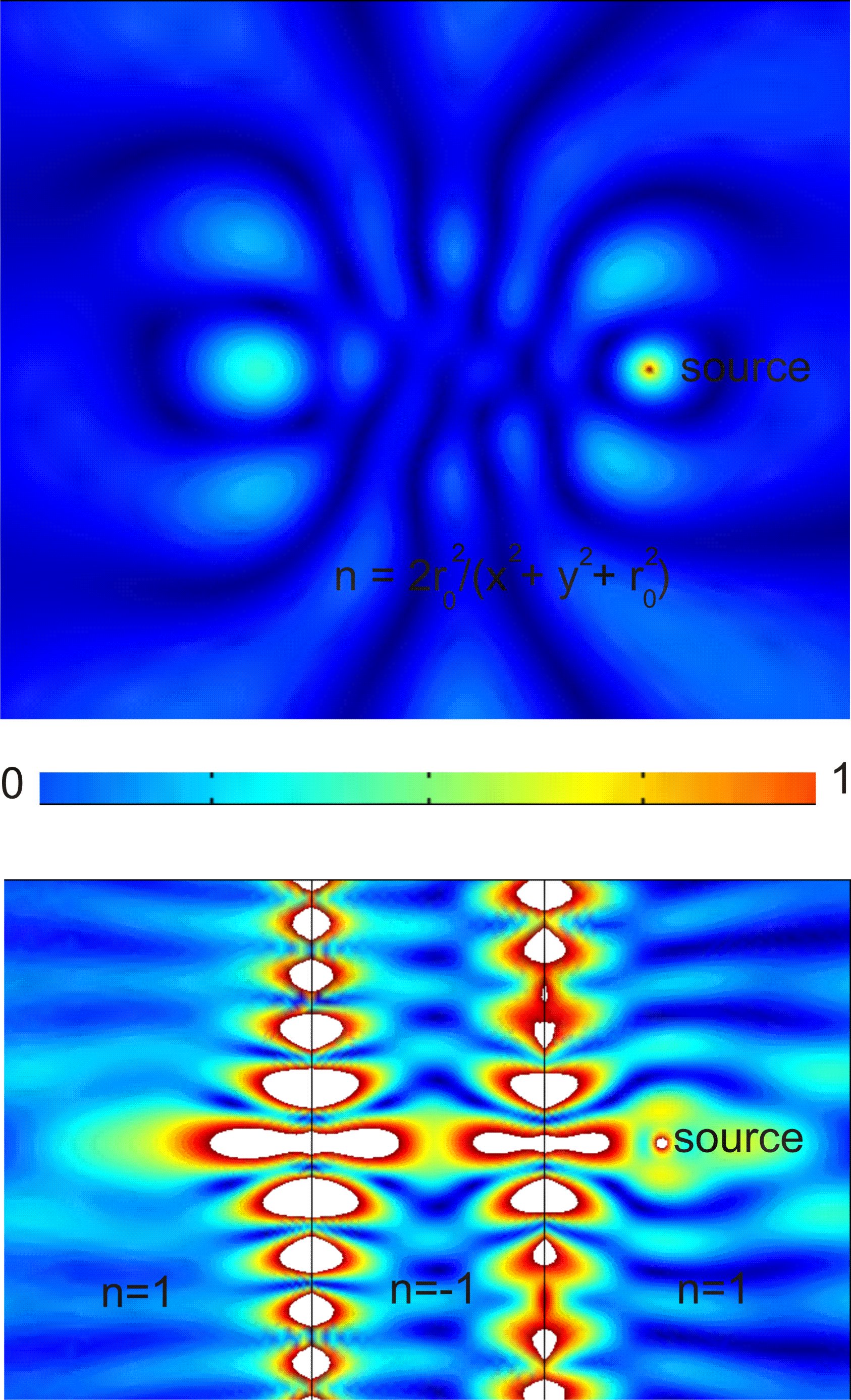}}
\mbox{}\vspace{-0.4cm}\caption{\em\small ~ 2D plot of the longitudinal electric field $\mid E_3\mid$:
(a) Lensing effect in the Veselago-Pendry silver slab lens
($\varepsilon=-1+i*0.4$); The harmonic line source (electric
current) at free space wavelength $\lambda_f=0.1$ is located a
distance $0.1$ away from the rightmost interface of the slab, and
the image appears a distance $0.1$ away from the leftmost interface;
(b) Lensing effect in the Maxwell fisheye associated with a virtual
sphere of radius $r_0=0.1$. The line source (electric current) at
free space wavelength $\lambda_f=0.1$ is located at point $(0.1,0)$
and the image appears at point $(-0.1,0)$ (note the astigmatism);
The image appears in the non-uniform refractive index in the Maxwell
fisheye whereas in the `perfect' lens, the image is formed in vacuum
but appears shifted with respect to the theoretical prediction.
} \label{fig6}
\end{figure}

\begin{figure}[h!]
\mbox{}\centerline{
\includegraphics[width=12cm,angle=-90]{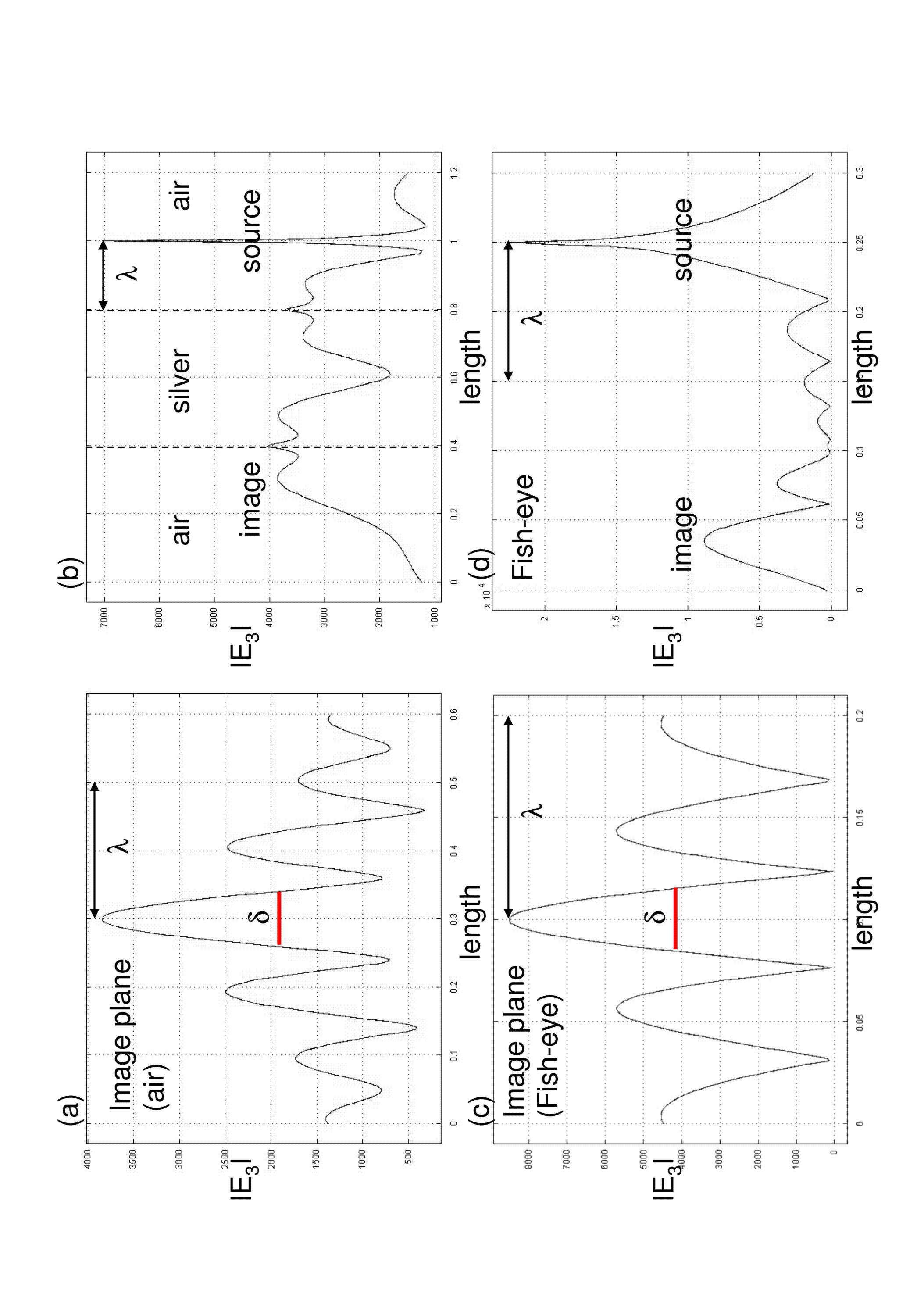}}
\mbox{}\caption{\em\small ~Upper panel: Modulus of the
longitudinal electric field $E_3$ radiated by a harmonic line source
(electric current) at free space wavelength $\lambda_f=0.2$, a
distance $0.2$ from a silver slab lens; (cf. lower panel of Fig.
\ref{fig6}); (a) Profile along the vertical direction in the plane
$x_1=0.3$ (whereas the theoretical prediction for the image plane is
$x_1=0.2$). The red line represents the resolution of the image:
full width at half maximum of the `image-point' is
$\delta\sim\lambda/3$, where $\lambda=\lambda_f$; (b) Profile of
$\mid E_3 \mid$ in the plane $x_2=0$ with a source at $x_1=1$, a
ghost image at $x_1\sim 0.6$ and an image at $x_1\sim 0.3$ (we note
the large amplitudes at the interfaces $x_1=0.4$ and $x_1=0.6$);
Lower panel: Modulus of the longitudinal electric field $E_3$
radiated by a harmonic line source (electric current) at free space
wavelength $\lambda=0.1$ in the Maxwell Fisheye associated with a
virtual sphere of radius $r_0=0.1$ (cf. upper panel of Fig.
\ref{fig6}); (c) Profile along the vertical direction in the image
plane $x_1=0.03$
The red line represents the resolution of the image: full width at
half maximum of the 'image-point' is $\delta\sim\lambda/3$, where
$\lambda<\lambda_f$ is the 'averaged wavelength' in the fisheye
computed from the source to the image planes; (d) Profile of $\mid
E_3 \mid$ in the plane $x_2=0$ with the source at $x_1=0.25$ and an
image at $x_1=0.03$.} \label{fig5}
\end{figure}

In the classical book by Born and Wolf \cite{born}, an optical
system is said to produce sharp imaging of an object-point ${\bf
x}_O$ onto an image point ${\bf x}_I$ when any ray trajectory
emitted from ${\bf x}_O$ through the optical system will pass
through ${\bf x}_I$ in an exact way. However, this definition might
be misleading, as the reconstruction of the image does not
necessarily involve the evanescent components of the source, thereby
not beating Rayleigh's diffraction limit.

Still in Born and Wolf \cite{born}, points ${\bf x}_O$ and ${\bf
x}_I$ are said to be perfect conjugates, and in our opinion this
definition is more accurate for the Maxwell fisheye. For instance,
the prolate ellipsoidal reflector is a well-known example of a
perfect imaging optical system, but only for points ${\bf x}_O$ and
${\bf x}_I$ coincident with the foci of the ellipsoid, and it does
not involve reconstruction of an image with both propagating and
evanescent components of the source (a necessary condition to beat
the diffraction limit). It seems therefore fair to call such a lens
a conjugate imaging optical system.

A class of conjugate imaging optical systems has been obtained in
curved lenses such as the generalized Maxwell fisheyes in
\cite{minano}. To construct these less than usual lenses, we
consider the following mapping from the sphere to the $xy$ Cartesian
plane
\begin{equation}
x={\left(\frac{1-\sin\theta}{\cos\theta}\right)}^{1/p}\cos(\frac{\phi}{p})
\; , \;
y={\left(\frac{1-\sin\theta}{\cos\theta}\right)}^{1/p}\sin(\frac{\phi}{p})
\; , \label{generalizedfisheye}
\end{equation}
where $p$ is an integer.

Noting that the radius of the sphere $S^2$ is $r_0=1/p$, it is
easily seen that
\begin{equation}
ds^2=\frac{{(d\theta)}^2+{(\cos\theta d\phi)}^2}{p^2}=n^2(dx^2+dy^2)
\; , \; n= \frac{2 (x^2+y^2)^{(p-1)/2}}{(x^2+y^2)^{p}+1} \; .
\label{infelementplanebis}
\end{equation}
Indeed, from (\ref{generalizedfisheye}) the total derivatives of $x$
and $y$ can be expressed as
\begin{equation}
\begin{array}{ll}
dx=\displaystyle{{\frac{1}{p}\left(\frac{1-\sin\theta}{\cos\theta}\right)}^{1/p}
\left(-\frac{\cos(\frac{\phi}{p})}{\cos(\theta)}d\theta-\sin(\frac{\phi}{p})d\phi\right)}
\; , \nonumber \\
dy=\displaystyle{{\frac{1}{p}\left(\frac{1-\sin\theta}{\cos\theta}\right)}^{1/p}
\left(-\frac{\sin(\frac{\phi}{p})}{\cos(\theta)}d\theta+\cos(\frac{\phi}{p})d\phi\right)}
\; ,
\end{array}
\label{proofgenemax1}
\end{equation}
so that
\begin{equation}
dx^2+dy^2=\displaystyle{{\frac{1}{p^2}\left(\frac{1-\sin\theta}{\cos\theta}\right)}^{2/p}
\left(\frac{1}{\cos^2(\theta)}d\theta^2+d\phi^2\right)} \; ,
\label{proofgenemax2}
\end{equation}
and (\ref{infelementplanebis}) follows by noting that
$n=\cos(\theta){\left(\frac{\cos\theta}{1-\sin\theta}\right)}^{1/p}$.

When $p=1$, (\ref{infelementplanebis}) reduces to
(\ref{infelementplane}), for which any object point ${\bf x}_O$ is
imaged in a point ${\bf x}_I$ such that ${\bf x}_O=-{\bf x}_I/{\bf
x}_I^2$, as can be checked from Fig. \ref{fig5}(d).

However, at least in the TM polarisation case, we will see in the
next subsection that such lenses cannot be called perfect in the
sense of Pendry's perfect lens, as the image does not contain
subwavelength components of the source: it is only in the ray optics
limit that the image of a point is a point, but the solution to the
Maxwell's equations will obey the Rayleigh diffraction limit.

If we now consider the case of the Pendry-Veselago slab lens, for
any pair of object points ${\bf x}_O$ and ${\bf x}_{O'}$, their
images fulfil $\mid {\bf x}_I-{\bf x}_{I'} \mid=\mid {\bf x}_O-{\bf
x}_{O'}\mid$.

One can actually prove the impossibility of sharply imaging two
object planes with non-unit magnification holds regardless of
distortion requirements \cite{born}. An image magnification $m$
means that $\mid {\bf x}_I-{\bf x}_{I'} \mid=m\mid {\bf x}_O-{\bf
x}_{O'}\mid$ and an example is provided by the homeopathy effect for
the perfect cylindrical lens \cite{milton1}. An interesting question
to ask is whether the magnified image is perfect, and this has been
actually shown by Pendry and Ramakrishna through a conformal mapping
from a periodic stack of heterogeneous anisotropic perfect lenses
onto the cylindrical homogeneous isotropic one \cite{sar1}.
Importantly, one of us actually studied  the electrostatic response
of a dielectric cylinder surrounded by a coating with negative
permittivity back in 1994 \cite{milton1}. This optical system also
acts as a perfect lens, which further magnifies the image, as
observed in \cite{sar1}.

In the same vein, it is legitimate to wonder whether
all-angle-negative refraction in photonic crystals \cite{anomalous}
refocusses a line source onto an image with a resolution lower or
equal to $\lambda/3$, since evanescent waves are not amplified in
this optical system. Merlin actually showed that the resolution of
the Veselago-Pendry lens scales as a logarithm of the absorption
within the lens \cite{merlin}. Merlin further demonstrated that the
absorption was linked to the spatial oscillations of surface waves
at the interfaces between complementary media, and this was further
numerically confirmed in general anisotropic heterogeneous
complementary media of the checkerboard type \cite{guenneau_ol05}.

The numerical simulation in the lower panel of Fig. \ref{fig6} for a
source radiating within the Maxwell fisheye, suggests that a perfect
image is formed within the non-uniform (positive) refractive index
of this most unusual lens. By comparison with the upper panel of
Fig. \ref{fig6}, whereby the negative refractive index within the
Pendry-Veselago slab lens
$n=-1+i*0.4$, it is indeed apparent that the image displays
comparable features. However, there is a major difference here: in
the former case, the highly heterogeneous region surrounding the
image contributes to its resolution, whereas in the latter case it
is fair to say that the image is subwavelength.

To understand the mechanism leading to the image formation within
the Maxwell fish eye, we look at the surface of the virtual sphere,
within which light rays propagate along geodesics, the great
circles. It is one of the remarkable properties of the stereographic
projection that circles on the sphere are transformed into circles
on the plane. From this we infer that in physical space, light goes
around in circles as well. The great circles originating from one
source point on the sphere meet again at the antipodal point. In the
stereographic projection, the image of the antipodal point is the
reflection of the source on a circle, the circle with the radius
$r_0$ of the sphere.

It is then tempting to claim that the source is perfectly imaged:
Maxwell's fish eye would make a perfect lens. However, this is quite
an usual instrument: both the source and the image are embedded in
the non-uniform refractive index profile of the fish eye. Clearly,
this optical system should display some astigmatism, which is clear
from the lower panel in Fig. \ref{fig6}: the optical system is not
symmetric about the optical axis. While this is also the case of a
cylindrical perfect lens \cite{milton2,pendry_jpc03,milton1}, we
note that astigmatism is observed even for rays from on-axis object
points in the fish eye. Hence, this lens is already less than
perfect. However, more importantly, the image appears in a highly
heterogeneous medium (the fish eye itself) and it is therefore not
simple to measure its resolution. The wavelength is obviously
compressed to a tiny space, but partly because of the large
refraction index.

In contrast, for the flat lens made with negative refractive index
material, both source and image are outside the device and in
perfect alignment about the optical axis. Note also that the lens
acts across some distance, unlike confocal microscopes or other
types of high-resolution optical system tomography \cite{hughes}.

\subsection{Leonhardt's proposal of a fisheye mirror in a nutshell}
The original fisheye, as first described by Maxwell, is a medium of
infinite extent. We therefore needed to cut off the fisheye with
some care as it focuses all rays from the object at the image, but
these include rays that propagate out to arbitrarily large radius.
If the fisheye is cut off carelessly, then the quality of the image
might deteriorate, but this would be a mere numerical artifact. We
therefore implemented PMLs within an annulus encompassing a disc
which is 10 wavelengths in diameter. Transverse magnetic waves are
then governed by (\ref{tm}) in the disc, and (\ref{tmpml}) in the
annulus and numerical results are reported in Fig. \ref{fig6} and
Fig. \ref{fig5}.

However, people might argue that adding PMLs within the model still
affects the numerical solution of the real open boundary problem.
Moreover, the refractive index for the fisheye varies between $n=2$
at the center and $n=1$ on the boundary of a disc of radius $r=r_0$
($r_0$ radius of the virtual sphere), and then decreases to zero at
infinity. This means that the speed of light outside the disk (for
$r>r_0$) is greater than $c$, speed of light in vacuum, and tends to
infinity when $r$ tends to infinity. A way of making a device with
the same imaging capability as the infinite fisheye and avoiding
such paradoxes is then to follow Leonhardt's recent proposal
\cite{fishulf} and put a mirror at $r=r_0$. This means we now keep
only (\ref{tm}), but importantly supply it with perfect electric
boundary conditions i.e. $E_3=0$ in this case of polarisation. The
numerical results are reported in Fig. \ref{fisheyemirror} and Fig.
\ref{fig9rev}.

To interpret results of Fig. \ref{fig5} and Fig. \ref{fig9rev},
namely the resolution of the image within a fisheye, it is necessary
to take into account the fact that the image forms in a complex
medium with varying refractive index greater than that of vacuum.
For this purpose, we adopt the following definition for the
resolution of an image within the Maxwell fisheye:

{\bf Definition:} {\it The resolution of the image in a
heterogeneous isotropic lens is given by the ratio of the full width
at half maximum of the image-point divided by the average of the
wavelength $\lambda$ between the source and image planes.}

If we adopt this definition, then the Maxwell fisheye has a
resolution of $\lambda/3$ in TM polarisation, which is just above
the Rayleigh diffraction limit. Indeed, one can see that in Fig.
\ref{fig5} and Fig. \ref{fig9rev} (upper panel) that the averaged
wavelength $\lambda$, is much smaller than the free space wavelength
(except outside the disc of radius $r_0$). If one would compute the
resolution with respect to free space wavelength, the image would be
considered as highly subwavelength, but this is in fact just the
classical effect of the wavelength contraction within any high
refractive index medium (an optical illusion!).

It is also interesting to look at the case of two line sources
radiating within the fisheye mirror lens. As we can see in the lower
panels of Fig. \ref{fisheyemirror} and fig. \ref{fig9rev}, there are
many secondary images appearing within the cavity. We also note that
the resolution of the images is only about $\lambda/2$, which could
attributed to the fact that the free space wavelength is twice as
small as in the upper panel. It is also possible that the larger the
number of sources the worse the resolution of their respective
images, but this can only be conjectured at this stage.

\begin{figure}[h!]\centerline{
\includegraphics[width=10cm]{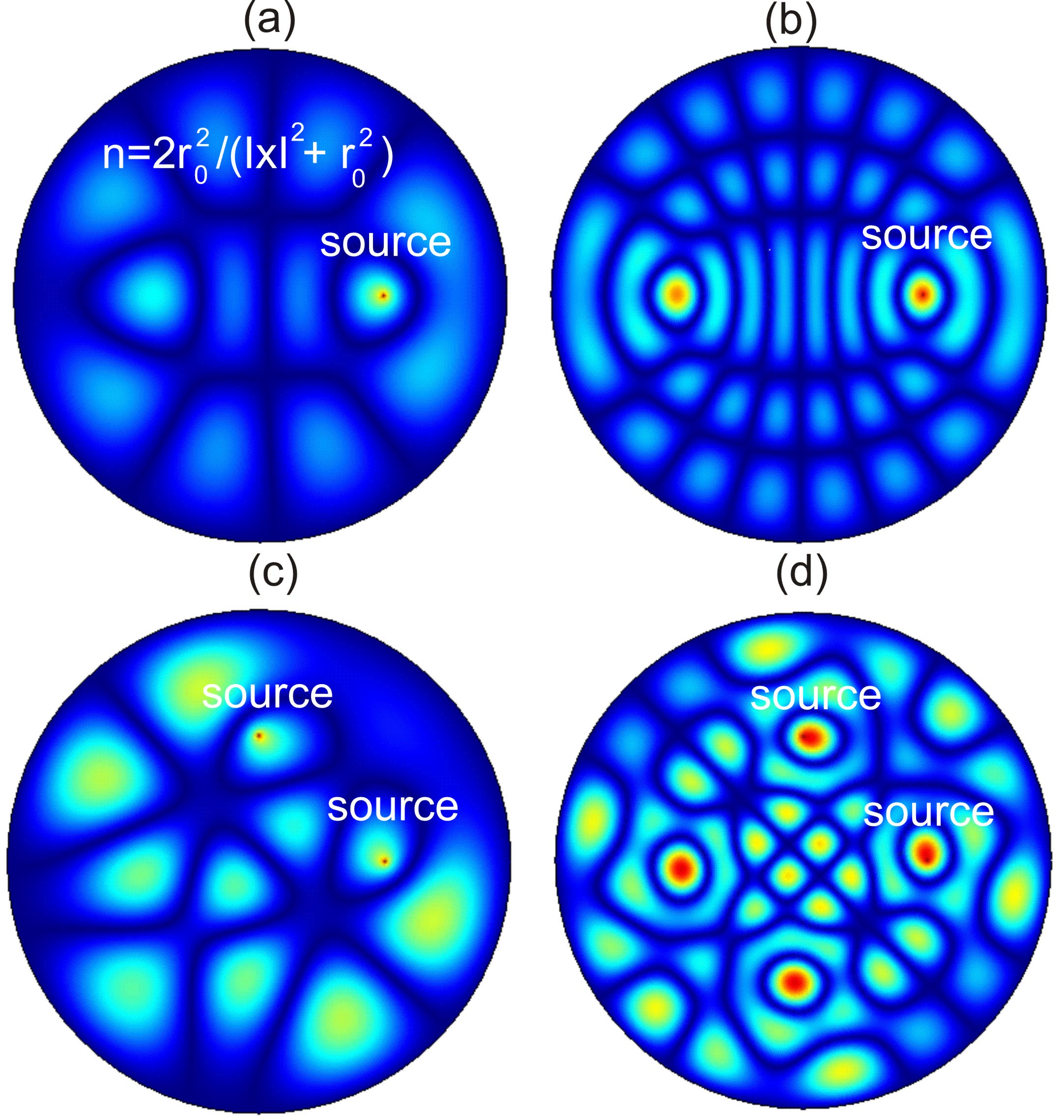}}
\caption{\em\small ~Harmonic line sources at free space wavelengths
$\lambda_f=0.2$ (a,c) and $\lambda_f=0.1$ (b,d) radiating within a
fisheye mirror of radius $r=0.2=r_0$ (with $r_0$ radius of the
virtual sphere) and refractive index defined by (\ref{infelement}):
(a,c) One source at point $(0.1,0)$; (b,d) Two sources at points
$(0.1,0)$ and $(0,0.1)$; We note the symmetry of the modulus of the
electric field $\mid E_3 \mid$ about the $y$-axis in (a,c) and about
$x=-y$ in (b,d).} \label{fisheyemirror}
\end{figure}

\begin{figure}[h!]\hspace{-.3cm}\mbox{}\centerline{
\includegraphics[width=12cm,angle=-90]{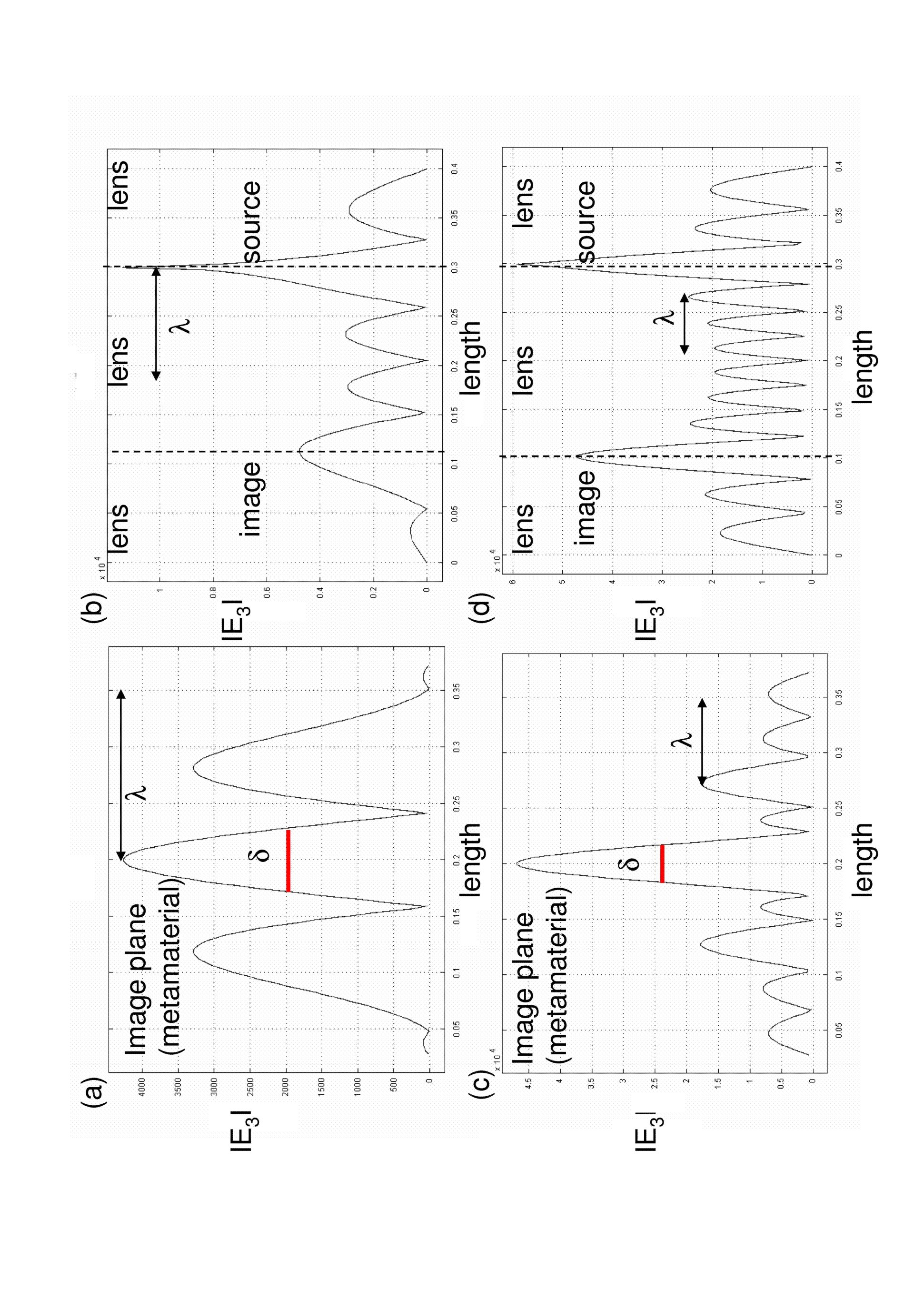}}
\mbox{}\caption{\em\small ~Upper panel: (a) Modulus of the
longitudinal electric field $E_3$ radiated by a harmonic line source
of free space wavelength $\lambda_f=0.2$ at point $(0.1,0)$ in the
fisheye mirror with radius $r=0.2=r_0$ (cf. upper-left panel of Fig.
\ref{fisheyemirror}) along the vertical direction in the plane
$x_1=0.12$. The red line represents the resolution of the image:
full width at half maximum of the 'image-point' is
$\delta\sim\lambda/3$ (with $\lambda<\lambda_f$); (b) Profile of
$\mid E_3 \mid$ in the plane $x_2=0$ with a source at $x_1=0.3$ and
an image at $x_1=0.12$; Lower panel: (c) Modulus of the longitudinal
electric field $E_3$ radiated by a line source of free space
wavelength $\lambda_f=0.1$ at points $(0.1,0)$ and $(0,0.1)$ along
the vertical direction in the image plane $x_1=-0.5$. The red line
represents the resolution of the image: full width at half maximum
of the 'image-point' is $\delta\sim\lambda/2$ (with
$\lambda<\lambda_f$); (d) Profile of $\mid E_3 \mid$ in the plane
$x_2=0$ with a source at $x_1=0.3$ and an image at $x_1=0.1$.}
\label{fig9rev}
\end{figure}

\subsection{Open questions on the closed fisheye cavity}
As mentioned in \cite{fishulf}, fisheye mirrors may find some
application in nanolithography, and are thus worthwhile
investigating both on the academic and technological sides.

\noindent In \cite{fishulf}, Leonhardt points out that {\it perfect
imaging does not occur for the TM polarization where the
magnetic-field vector H points orthogonal to the plane}. However,
the terminology used in Leonardt's paper refers to (\ref{te}) as the
TM case, whereas we refer to this polarisation as transverse
electric (TE). We therefore need to study the other polarization
case to backup this statement.

\begin{figure}[h!]\centerline{
\includegraphics[width=10cm]{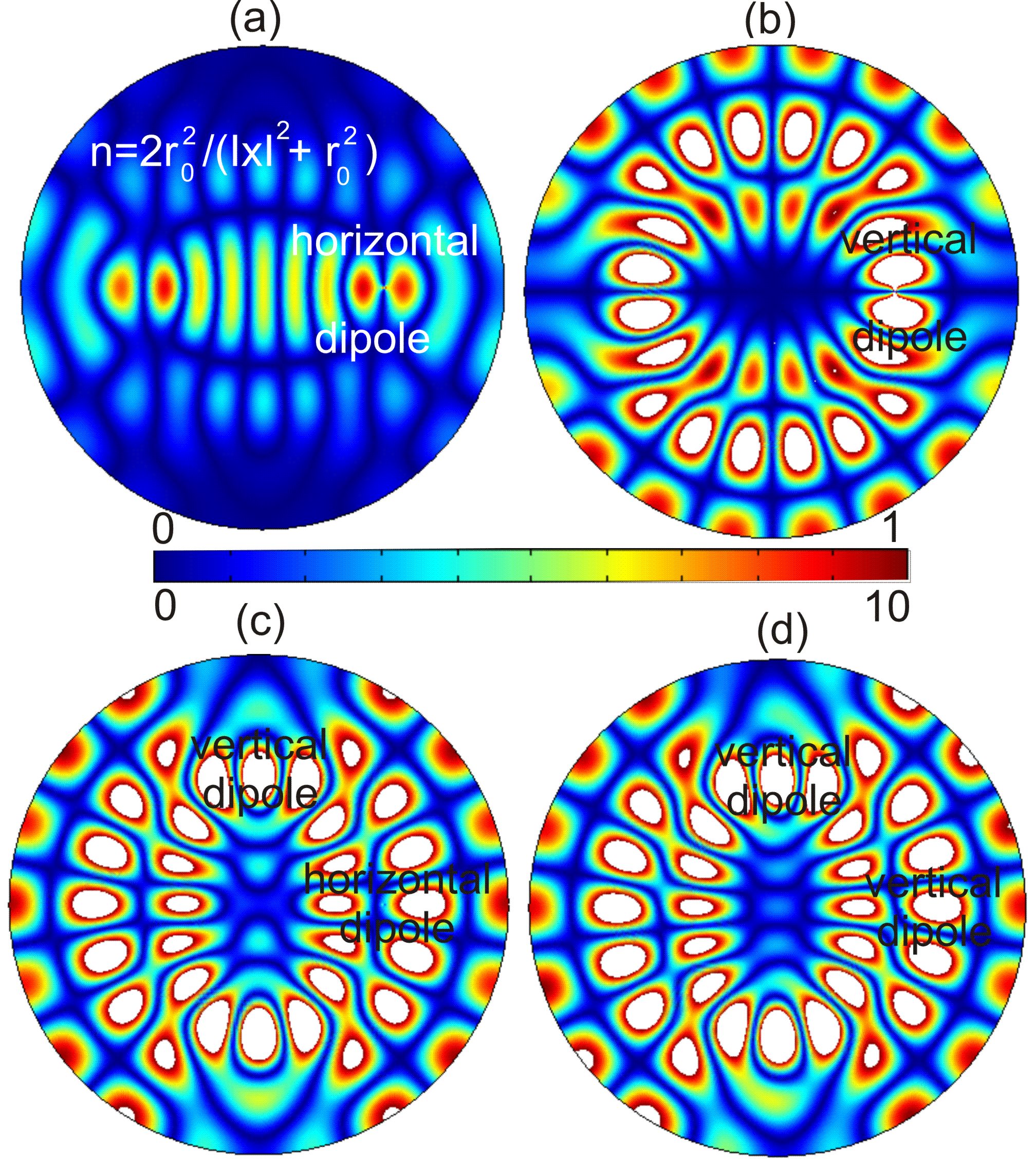}}
\caption{\em\small ~Harmonic dipole sources at free space wavelength
$\lambda_f=0.1$ radiating within a fisheye mirror of radius
$r=0.2=r_0$ (with $r_0$ radius of the virtual sphere) and refractive
index defined by (\ref{infelement}): (a,b) One dipole at point
$(0.1,0)$ oriented along the horizontal (a) and vertical (b) axes;
(b,d) Two dipole sources at points $(0.1,0)$ and $(0,0.1)$ oriented
along the horizontal and vertical axes; We note the symmetry of the
modulus of the magnetic field $\mid H_3 \mid$ about the $y$-axis in
(a,b,c), and the huge enhancement of the field in (c,d) requiring a
color scale ten times larger.} \label{fisheyemirrorte}
\end{figure}

\begin{figure}[h!]\mbox{}\centerline{
\includegraphics[width=12cm,angle=-90]{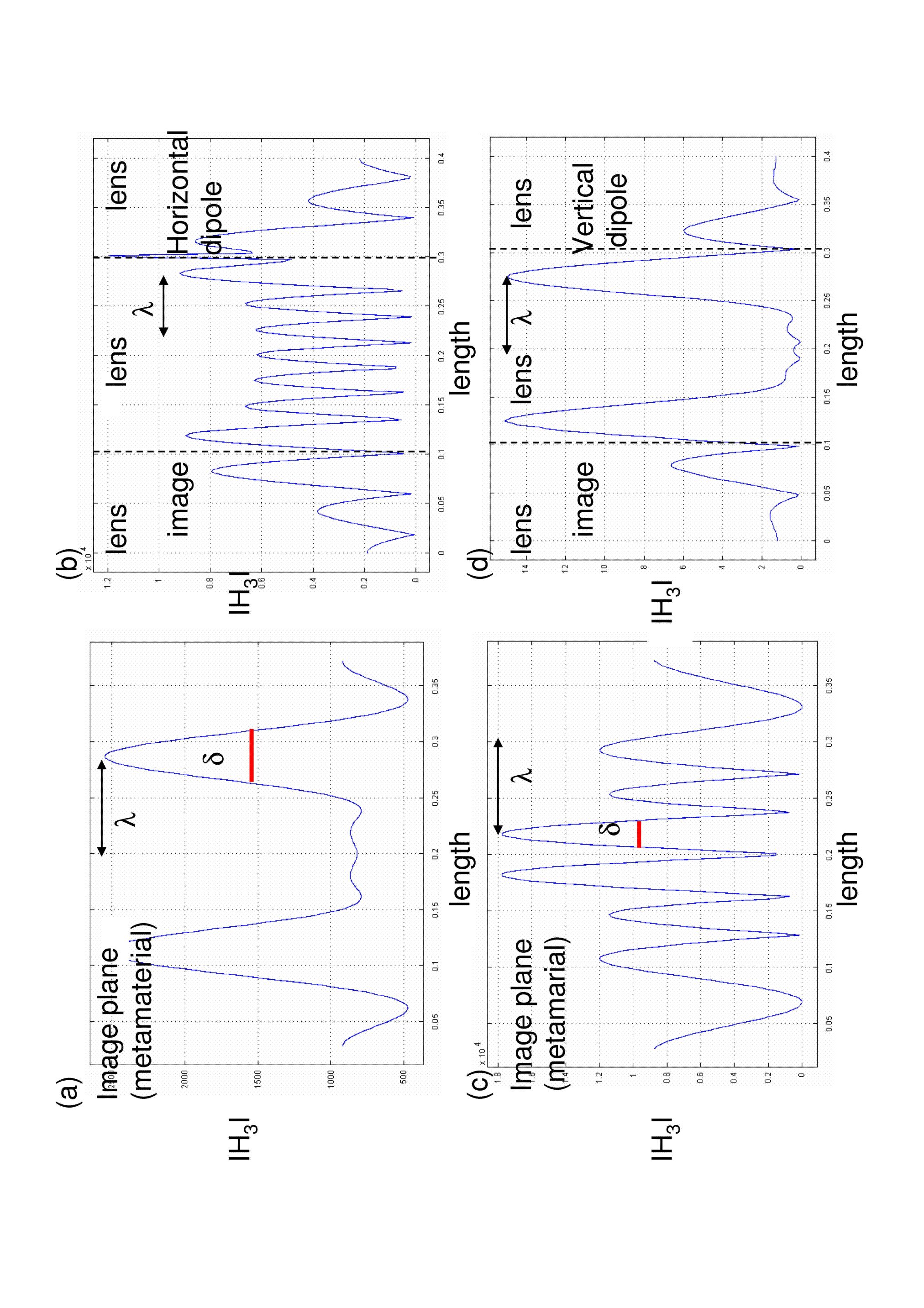}}
\mbox{}\caption{\em\small ~ Upper panel: (a) Modulus of the
longitudinal magnetic field $H_3$ radiated by a magnetic dipole
oriented along $x_1$ with free space wavelength $\lambda_f=0.1$ at
point $(0.1,0)$ in the fisheye mirror with radius $r=0.2=r_0$ (cf.
upper-left panel of Fig. \ref{fisheyemirrorte}) along the vertical
direction in the plane $x_1=0.1$. The red line represents the
resolution of the image: full width at half maximum of the
'image-point' is $\delta\sim\lambda/4$ (with $\lambda<\lambda_f$);
(b) Profile of $\mid H_3 \mid$ in the plane $x_2=0$ with a dipole
source at $x_1=0.3$ and an image at $x_1=0.1$; Lower panel: (c)
Modulus of the longitudinal magnetic field $H_3$ radiated by a
dipole source oriented along $x_2$ with free space wavelength
$\lambda_f=0.1$ at point $(0.1,0)$ in the image plane $x_1=0.1$. The
red line represents the resolution of the image: full width at half
maximum of the 'image-point' is $\delta\sim\lambda/2$ (with
$\lambda<\lambda_f$); (d) Profile of $\mid H_3 \mid$ in the plane
$x_2=0$ with a source at $x_1=0.3$ and an image at $x_1=0.1$.}
\label{fig9rev11}
\end{figure}

As mentioned earlier, this case represents a challenge in terms of
modelling, because of the presence of a magnetic source. In order to
provide the reader with an idea of what is going on in this case of
polarisation, we decided to approximate the source by a set of two
closely located line sources, with opposite charges, and this indeed
result in a magnetic dipole, as can be seen in Figure
\ref{fisheyemirrorte}. Here, the two point sources are located a
distance $0.002$ apart while the free space wavelength is
$\lambda=0.1$. We set some infinite conducting boundary conditions
i.e. $\partial H_3/\partial n=0$ at $r=r_0$. We observe that the
orientation of the moment of the magnetic dipole (here, along the
horizontal and vertical axes) indeed matters for the reconstruction
of the image, as can be seen in Fig. \ref{fig9rev11}. In this case
of polarisation, we have to adapt our definition of image resolution
to a dipole, and it seems natural to now look at the ratio of the
full width at half maximum of the dipole image points divided by the
average of the wavelength $\lambda$ between the source and image
planes, which corresponds to the red segment in Fig.
\ref{fig9rev11}.

Actually, we find that the image resolution varies between
$\lambda/2$ and $\lambda/4$, the latter being subwavelength. We
checked that when we only consider a line source (a magnetic
monopole), the resolution of the image does not exceed $\lambda/2$,
and lies therefore clearly within the Rayleigh diffraction limit.
The statement of Leonhardt is therefore fully consistent with our
findings in the case of a magnetic monopole, but the answer seems
less conclusive for a magnetic dipole. However, we are not yet in a
position to either prove or disprove Leonhardt's statement that in
the other case of polarisation (TM) {\it the geometry of light
established by Maxwell's fisheye is not restricted to rays, but
extends to waves, which may explain why waves are as perfectly
imaged as rays.} More precisely, according to Leohnardt, for a 2D
fisheye formed by an electric permittivity with trivial magnetic
permeability, only for the TM polarization (using our terminology
i.e. for (\ref{tm})) is wave propagation exactly equivalent to
propagation on a sphere and the image of a point is a point. Our
numerical results suggest that the resolution of the image still
depends upon the material constituents and with our definition is
about one third of the wavelength. Moreover, Leonhardt's proposal of
a 3D fisheye that would image all waves perfectly refers to a medium
that exactly corresponds to propagation on a hypersphere, which
requires a Maxwell fisheye with equal permittivity and permeability
given by Maxwell's refractive index profile.

\noindent In the course of this work, we came across other
interesting features on the fisheye, apparently not discussed in the
literature, which we now discuss.
\section{Cross-over ray trajectories and multiple images in the Fisheye}
There is yet another issue to address for the Maxwell fisheye: if
one draws some ray trajectories emanating from two sources on the
virtual sphere as in Fig. \ref{fig2}, it appears that there are many
images resulting from the many intersection of geodesics (great
circles). It looks as if the set of images is a dense subset of the
surface of the sphere i.e. there are an infinite number of images.
This is a very surprising feature whereby one source gives rise to
one and only one image whereas two sources give rise to an infinite
number of images. This ray picture is confirmed by numerical
simulations in Fig. \ref{fig3} which clearly show that there are
more and more images when the wavelength is smaller and smaller.
When we move to higher frequencies, we observe even more images.
However, when we refine the mesh solutions become less stable for
higher frequencies. Our computations suggest that we would observe
many more images if we could refine the mesh even further within our
finite element model. The solution of the real problem lies maybe
somewhere between the ray optics picture indicating an infinite
number of cross-over ray trajectories on the sphere and the wave
picture which suggest that the number of images depends upon the
wavelength, and somewhat on the mesh used in the numerics. We
believe this is a very interesting electromagnetic paradigm well
worth investigating with more analytical tools to check if there are
no numerical artifacts behind the wave pictures depicted here.

\begin{figure}[h!]\centerline{
\includegraphics[width=10cm]{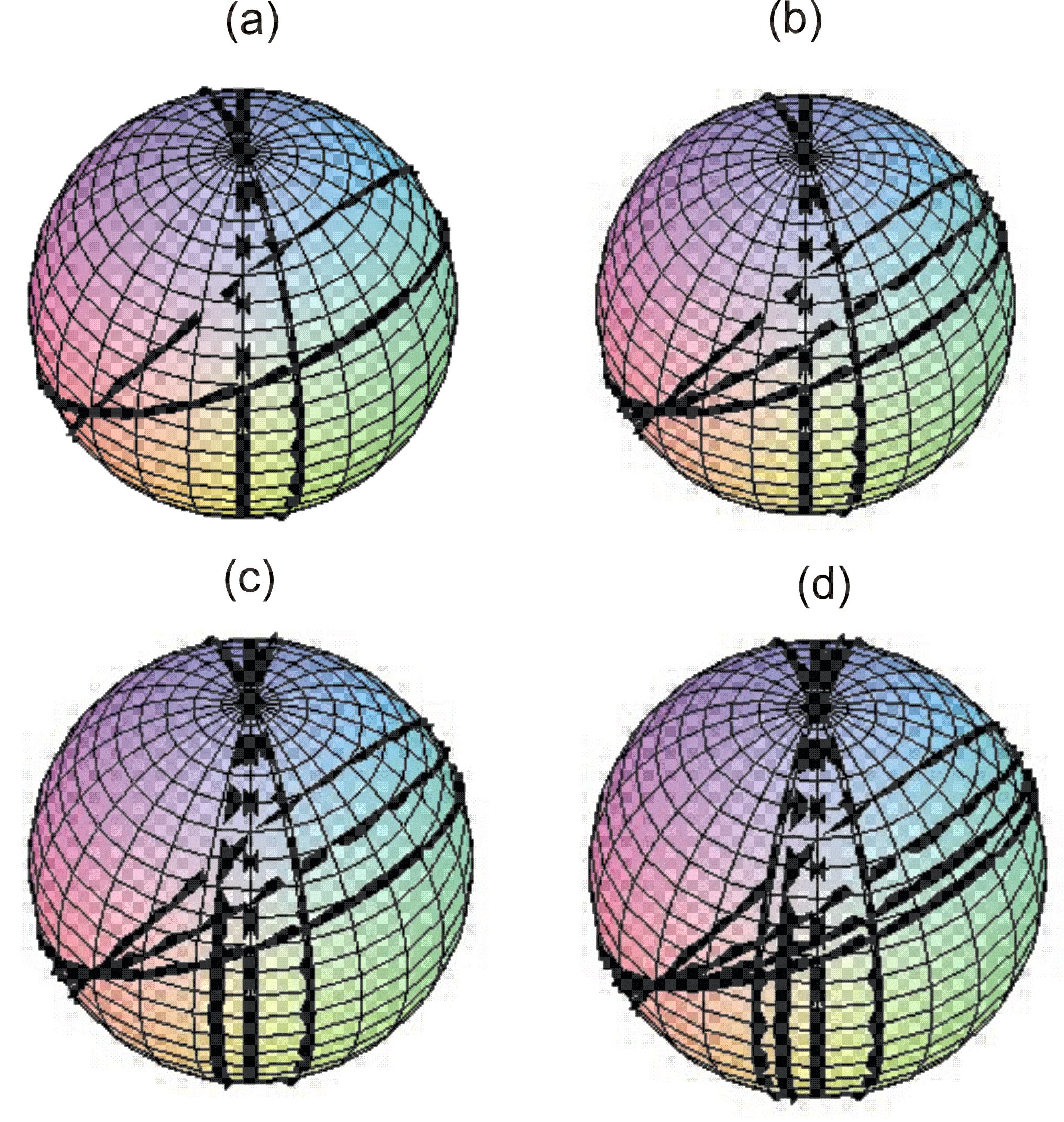}}
\caption{\em\small ~ Ray trajectories for light radiating by two sources points
at the surface of the virtual sphere with (a) 2 rays radiated by
each source; (b) 3 rays radiated by one source and 2 rays by the
other; (c) 3 rays radiated by each source; (d) 4 rays radiated by
each source; Note that the more rays emanating from the two sources,
the more intersections between ray trajectories and therefore the
more images.} \label{fig2}
\end{figure}

\begin{figure}[h!]\centerline{
\includegraphics[width=10cm]{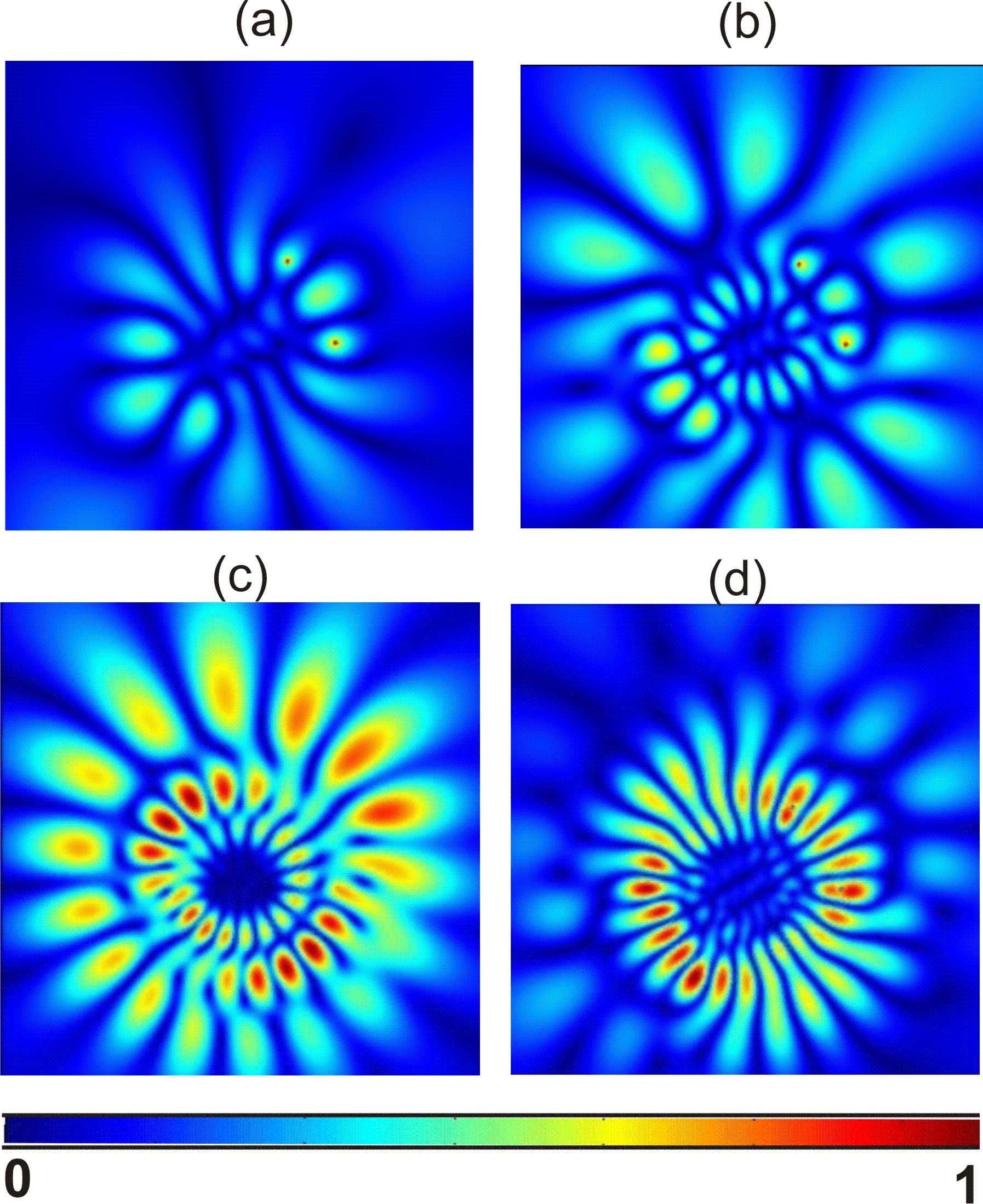}}
\caption{\em\small ~ Mirages effects: two sources in the Maxwell fisheye produce
multiple images as shown in Fig. \ref{fig2}; Modulus of $E_3$ two
harmonic line sources located at $(1.5,0)$ and $(0,1.5)$ radiating
at free space wavelengths (a) $\lambda_f=0.4$; (b) $\lambda_f=0.27$;
(c) $\lambda_f=0.24$; (d) $\lambda_f=0.2$.} \label{fig3}
\end{figure}

There is an interesting analogy between the ray construction in
optics and electromagnetism and the method of images in
electrostatics. The former is sometimes evident in expansions like
the Airy series, which treats multiple reflections between
interfaces as a sum over rays. The latter can be used to form an
imaging series between two dielectric interfaces when solving
Laplace's equation. The interesting issue which arises then
\cite{rossref} is the convergence of the imaging series, which
indeed is handled in electrostatics by the method of analytic
continuation. The resulting expression is a function called the
dilogarithm, which has a branch cut along the negative real axis of
dielectric permittivity. This analogy suggests there may well be
difficulties with the ray picture of electromagnetism when negative
permittivities and permeabilities are present, and that field
solutions embodying analytic functions embody the resolution of
these difficulties.

\section{Coming to the point in the Eaton lens}
From the previous analysis, we can see that the image resolution we
obtained in the Maxwell fisheye and the silver slab lens is about
one third of the wavelength in vacuum and cannot be therefore
classified as deeply subwavelength. However, in the former lens,
both source and image lie within a heterogeneous medium, whereas in
the latter, they are located in air (and far from the interface
metamaterial-air). Therefore, we are facing a dilemma as none of
these lenses clearly meets the requirement for a perfect lens (i.e.
the image beats the diffraction limit): they are borderline.
However, in the case of the silver slab lens, we know that the
resolution is limited by the absorption in the negative refractive
index medium, and the image is also formed in vacuum, and far from
the interface between metamaterial and air (confocal microscopes can
indeed reach a resolution of $\lambda/20$, however only in the
intense near field limit). From the 2005 experiment of Zhang's team,
a resolution of $\lambda/6$ has only been achieved with a thin film
of silver, and alternating layers of silver and dielectric might
further improve this resolution \cite{sar1}.

We thus proceed with a last example of conjugate optical imaging
system, known as the Eaton lens \cite{luneburg}. In that case, the
refractive index is defined as follows:
\begin{equation}
n= \left\{
\begin{array}{lll}
1 & \; , \; \hbox{ for } \sqrt{x^2+y^2} <1 \; , \nonumber \\
\sqrt{\frac{2}{\sqrt{x^2+y^2}}-1} & \; , \; \hbox{ for } 1 \leq
\sqrt{x^2+y^2} < 2 \; .
\end{array}
\right. \label{eatonlens}
\end{equation}
The refractive index is not defined outside the disc
$\sqrt{x^2+y^2}\leq 2$. On the boundary of this disc, perfect
conducting boundary conditions are imposed i.e. the lens is
surrounded by a mirror.

\noindent We note that the refractive index in the ring $1 \leq
\sqrt{x^2+y^2} < 2$ is lower than that of the unit disc which is
filled with air. This explains why the averaged wavelength $\lambda$
between the source and image planes appears to be smaller than the
wavelength in the image plane (filled with $n<1$) in Fig.
\ref{fig7bis}.

\noindent Ray trajectories inside the unit disc are obviously
straight lines ($n=1$) and since the refractive index is always
symmetric about $x=0$, it is easily seen that any source has a
mirror image such that ${\bf x}_I=-{\bf x}_0$.

\noindent This prediction of the ray optics limit corresponds to
what we obtain with the full wave solution, see Fig. \ref{fig7}.
However, the image resolution we computed from Figs. \ref{fig7} and
\ref{fig7bis} is about half of the averaged wavelength. It seems
therefore fair to say that while the geodesics in the Maxwell
fisheye and the Eaton lens suggest focusing of a source comes to an
image point, this theoretical prediction breaks down when look for
subwavelength imaging by solving the Maxwell equations. Negative
refraction makes a perfect lens, but the Maxwell fisheye allows for
conjugate images in the ray optics limit.

\begin{figure}[h!]\centerline{
\includegraphics[width=10cm]{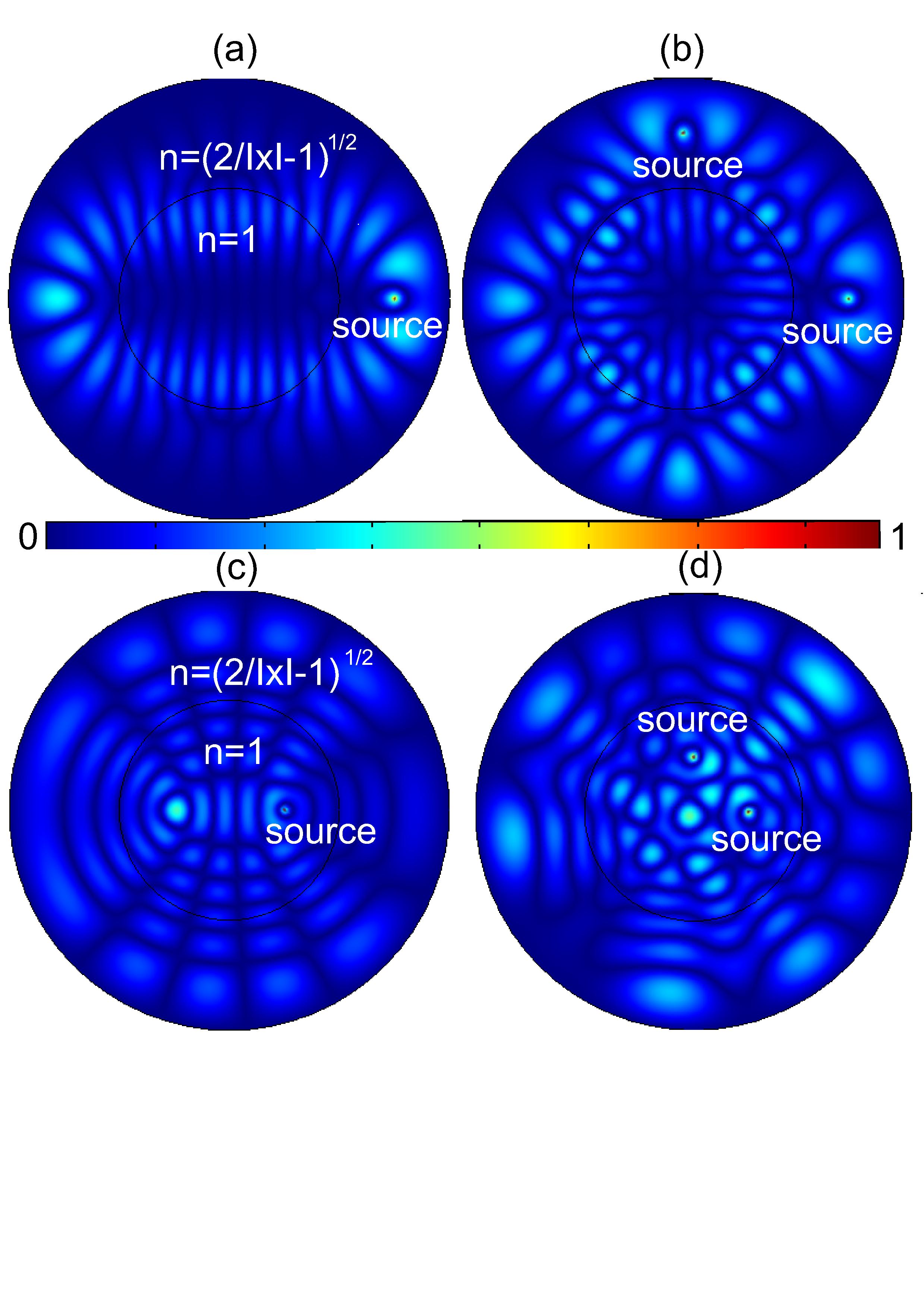}}
\vspace{-4cm}\mbox{}
\caption{\em\small ~ Harmonic line sources at wavelength $\lambda=0.4$ radiating
within an Eaton lens of radius $0.2$ and refractive index defined by
(\ref{eatonlens}): (a) One source at point $(1.5,0)$; (b) Two
sources at points $(1.5,0)$ and $(0,1.5)$; (c) One source at point
$(0.5,0)$; (d) Two sources at points $(0.5,0)$ and $(0,0.5)$.}
\label{fig7}
\end{figure}

\begin{figure}[h!]\mbox{}\centerline{
\includegraphics[width=12cm,angle=-90]{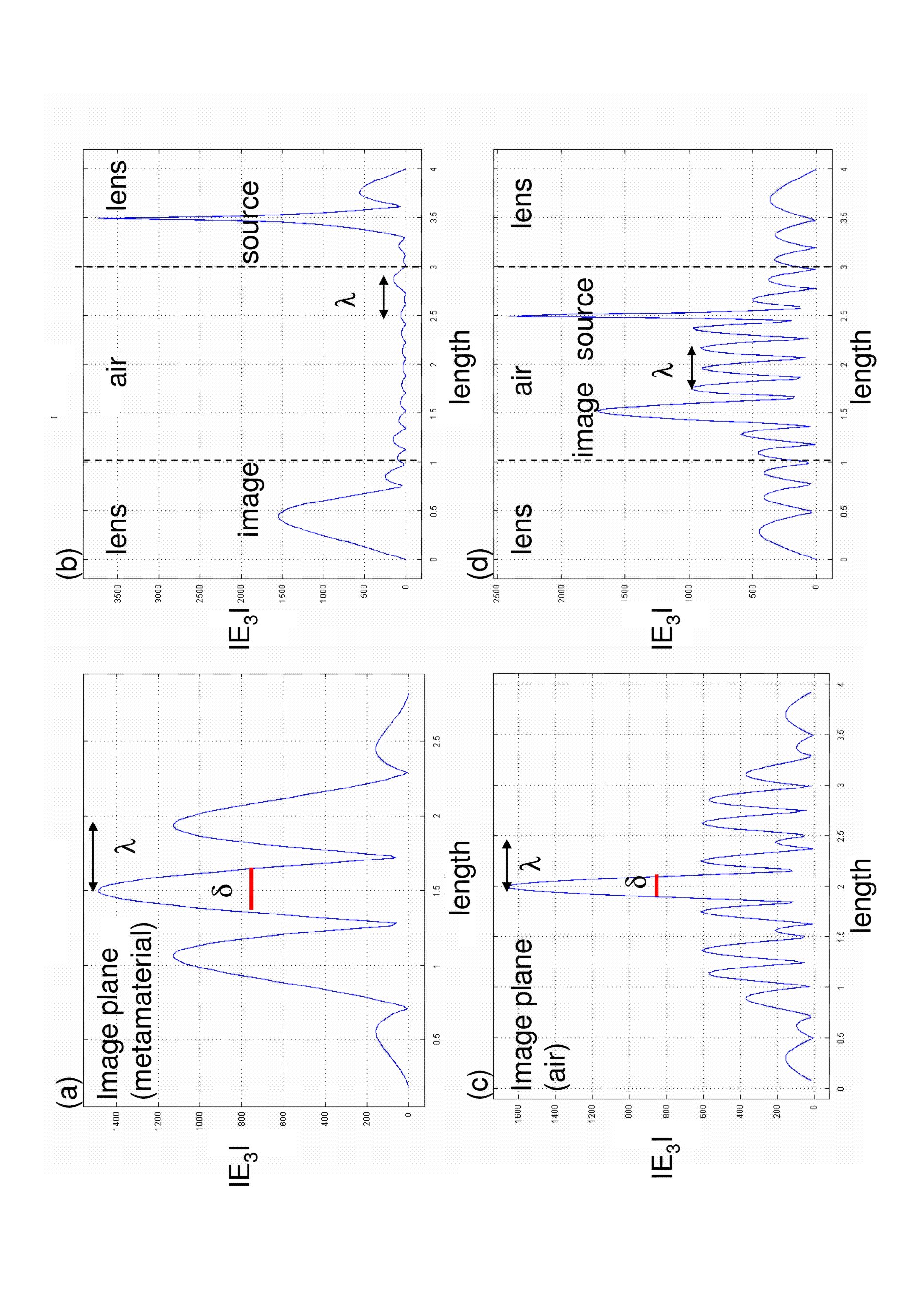}}
\mbox{}\caption{\em\small ~Upper panel: (a) Modulus of the
longitudinal electric field $E_3$ radiated by a line source of free
space wavelength $\lambda_f=0.2$ at point $(1.5,0)$ in the Eaton
lens (cf. upper-left panel of Fig. \ref{fig7}) along the vertical
direction in the plane $x_1=-1.5$. The red line represents the
resolution of the image: full width at half maximum of the
'image-point'; (b) Profile of $\mid E_3 \mid$ in the plane $x_2=0$;
Lower panel: (c) Modulus of the longitudinal electric field $E_3$
radiated by a line source of free space wavelength $\lambda_f=0.1$
at point $(0.5,0)$ (in air) in the Eaton lens (cf. lower-left panel
of Fig. \ref{fig7}) along the vertical direction in the image plane
$x_1=-0.5$. The red line represents the resolution of the image:
full width at half maximum of the 'image-point'; (d) Profile of
$\mid E_3 \mid$ in the plane $x_2=0$.} \label{fig7bis}
\end{figure}

\section{Concluding remarks}
In conclusion, we have presented a comprehensive variational
analysis of the Fermat's principle leading to the equation for
geodesics. We further investigated the design of the Maxwell fisheye
and the Pendry-Veselago slab lens using transformation optics. This
enabled us to notice three main differences between these optical
systems: the former displays some astigmatism while the latter does
not; the former focusses a source onto an image in a heterogeneous
medium while the latter displays an image in vacuum; the former
shows an infinity of secondary images in the case of two sources
while the latter only shows two images. We numerically noted that
the Maxwell fisheye and a silver slab lens both display a resolution
of $\lambda/3$ in the transverse magnetic case (electric field
pointing orthogonal to the plane), which can be considered
borderline for super-resolution. In the transverse electric
polarisation, the resolution of the image (a dipole) varies between
$\lambda/2$ and $\lambda/4$ depending upon the orientation of the
magnetic moment of the dipole source. However, when we looked at the
Eaton lens, the fact that its resolution does not exceed $\lambda/2$
confirms us in our opinion that while a negative refractive index
can lead to subwavelength imaging, the limit of image resolution
(and its very definition) within a Maxwell fisheye is far from being
obvious.

Forbes and Wallace have noted back in 1995, that there is a fine
line between the system performance displayed by a lens in
geometrical optics and the resolution which can effectively be
attained from full wave computations \cite{forbes}. However,
Leonhardt has mathematically shown \cite{fishulf} that for a Maxwell
fisheye with a mirror at $r=r_0$ that the analytical form of the
Green's function provides the right logarithmic singularity to
expect a perfect image in the transverse magnetic case (electric
field pointing orthogonal to the plane), unlike for the transverse
electric case. Our numerical results are thus less contrasted.
Leonhardt further claims in \cite{fishulf} that a three-dimensional
version of the Maxwell fisheye (i.e. constructed from a 4D
hypersphere) would make a perfect lens without negative refraction.
While this proposal seems to be a promising route towards the design
of new transformation based optical devices, the numerical study
required to prove the subwavelength features of this less than
ordinary lens lies beyond the scope of the present paper.

In our opinion, while the Maxwell fisheye is in itself a fascinating
optical system, it rather forms conjugate images than perfect
images, at least in two dimensions \cite{fishulf}. We agree that all
rays emanating from a source converge to an image, thereby forming a
sharp image in the ray optics limit. However, negative refraction is
a prerequisite for existence of plasmons enabling the amplification
of the evanescent component of the near field radiated by the
source, thereby clearly making the image subwavelength in the
absence of strong absorption or losses. While it might be too much
to ask a thick slab of silver to act as a perfect lens for some
optical wavelength, an alternation of thin silver films and
dielectrics as first proposed in \cite{sar3} might work just fine.
It is also possible to surround the slab of silver with air on one
side and other media such as glass or GaAs on the other to make an
asymmetric lossy near-perfect lens an alternation of thin silver
films and dielectrics as first proposed in might work just fine
\cite{sar1}.

The renewed interest in classical lenses such as the Maxwell fisheye
is fueled by the original viewpoint of transformation optics, which
is a burgeoning field still in its infancy. Analogies with general
relativity whereby the interplay of gravitational waves with
space-time metric becomes more apparent when equations are written
in covariant forms lead to fascinating devices based on projections
from upper-dimensional spaces. For instance, the design of a 3D
non-singular cloak or a 3D Maxwell fisheye requires analysis of
geodesics on a hypersphere \cite{ulflast}. These represent
unprecedent challenges for computational electromagnetism associated
with physical paradigms of interest for a general audience.
Correspondences with other areas of physics where governing
equations are transformation invariant such as quantum physics,
elastodynamics, fluid dynamics have already been noticed
\cite{milton,prl2008,zhang08,mic2009,prbihar}. Needless to say that
new discoveries are on their way and anyone wishing to take part in
this cutting-edge research might become pioneer of a new field of
science in his/her own right.

\section*{Acknowledgements}
AD and SG acknowledge funding from the Engineering and Physical
Sciences Research Council grant EPF/027125/1. RCM acknowledges
support from the Australian Research Council Discovery Grants
Scheme.

\label{lastpage}

\end{document}